\documentclass[twocolumn,tighten]{aastex6}
\usepackage{mathptmx}
\usepackage[latin9]{inputenc}
\usepackage{booktabs}
\usepackage{amstext}
\usepackage{graphicx}

\makeatletter

\providecommand{\tabularnewline}{\\}

\newcommand{\chandra}{{\it Chandra}}

\makeatother

\begin{document}

\title{Stripping of the Hot Gas Halos in Member Galaxies of Abell 1795}

\author{Cory R. Wagner}

\affil{Department of Physics, Engineering Physics \& Astronomy, Queen's
University, 64 Bader Lane, Kingston, Ontario, K7L 3N6, Canada}

\email{cory.wagner@queensu.ca}

\author{Michael McDonald}

\affil{Kavli Institute for Astrophysics and Space Research, Massachusetts
Institute of Technology, 77 Massachusetts Avenue, Cambridge, MA, 02139,
USA}

\author{St\'{e}phane Courteau}

\affil{Department of Physics, Engineering Physics \& Astronomy, Queen's
University, 64 Bader Lane, Kingston, Ontario, K7L 3N6, Canada}
\begin{abstract}
The nearby cluster Abell 1795 is used as a testbed to examine whether
hot gas in cluster galaxies is stripped by the ram pressure of the
intracluster medium (ICM). The expected X-ray emission in and around
Abell 1795 galaxies is likely dominated by the ICM, low-mass X-ray
binaries, active galactic nuclei, and hot gas halos. In order to constrain
these components, we use archival \textit{Chandra X-ray Observatory}
and Sloan Digital Sky Survey (SDSS) observations of Abell 1795 and
identify 58 massive ($M_{\star}>10^{10}\,M_{\odot}$) spectroscopic
cluster members within $5\arcmin$ of the \chandra\ optical axis.
X-ray images at 0.5\textendash 1.5 keV and 4\textendash 8 keV were
created for each cluster member and then stacked into two clustercentric
radius bins: inner ($0.25<R_{\mathrm{clust}}/R_{500}<1$) and outer
($1<R_{\mathrm{clust}}/R_{500}<2.5$). Surface brightness profiles
of inner and outer cluster members are fit using Markov chain Monte
Carlo sampling in order to generate model parameters and measure the
0.5\textendash 1.5 keV luminosities of each model component. Leveraging
effective total \chandra\ exposure times of 3.4 and 1.7 Msec for
inner and outer cluster members, respectively, we report the detection
of hot gas halos, in a statistical sense, around outer cluster members.
Outer members have 0.5\textendash 1.5 keV hot halo luminosities ($L_{X}=\left(8.1_{-3.5}^{+5}\right)\times10^{39}\,\mathrm{erg\,s^{-1}}$)
that are six times larger than the upper limit for inner cluster members
($L_{X}<1.3\times10^{39}\,\mathrm{erg\,s^{-1}}$). This result suggests
that the ICM is removing hot gas from the halos of Abell 1795 members
as they fall into the cluster.
\end{abstract}

\keywords{galaxies: clusters: general \textemdash{} galaxies: evolution \textemdash{}
X-rays: galaxies}

\section{Introduction}

\label{sec:intro}

The correlation between a galaxy's star formation activity and its
environment and look-back time has been known for quite some time.
Relative to the low-density (field) environment, the star-formation
activity of galaxies in dense galaxy clusters is much lower \citep{balogh1997}.
Indeed, local clusters are predominantly populated by galaxies that
are no longer forming stars \citep[they are said to have been quenched or to be quiescent;][]{chung2011}.
Furthermore, observations of distant clusters reveal star-formation
activity that is higher than what is found locally \citep{tran2010},
and in some cases as high as in the field \citep{brodwin2013,alberts2014}.
Clearly, environment has played a significant role in quenching (turning
off) the star formation of galaxies.

Galaxies become quiescent through the depletion of the cold gas necessary
for the formation of new stars. Several mechanisms, which act over
very different timescales, have been proposed to explain the observed
quenching of star formation in cluster galaxies over time. Active
galactic nucleus (AGN) feedback is a short-timescale quenching process
\citep[$\sim$100 Myr;][]{dimatteo2005} that can heat and expel cold
interstellar gas \citep{hopkins2006}. Ram-pressure stripping \citep[RPS;][]{gunn1972}
will also quickly \citep[$\leq$1 Gyr;][]{quilis2000} remove a galaxy's
neutral gas due to the pressure exerted as it moves through the diffuse
intracluster medium (ICM). This differs from the much slower process
of strangulation \citep[$>$1 Gyr;][]{larson1980}, by which the loosely-bound
hot halo of a galaxy can be stripped by the same ram pressure during
infall into the cluster. The in situ cold gas, however, is not removed
and the galaxy can continue to form new stars until this supply is
depleted, which can take several Gyr.

While the above quenching mechanisms all act on some level in clusters,
recent observational studies provide indirect support for strangulation
being the dominant process. These lines of evidence include the elevated
stellar metallicities in cluster galaxies \citep{peng2015}, consistent
with long quenching timescales, and stellar populations that indicate
long periods of persistent low-level (compared to the field) star
formation \citep{paccagnella2016}. However, a more direct approach
can be taken: measuring the gas content of galaxies as a function
of environment. From an X-ray perspective, the observational signature
of strangulation would be a transition around the virial radius of
galaxy clusters from gas-rich hot halos at large clustercentric radius
to gas-poor hot halos in the inner cluster.

Ideally, one would measure the hot halo luminosity for all cluster
galaxies as a function of clustercentric radius. However, the electron
density of individual halos is $n_{e}\sim10^{-4}\,\mathrm{cm}^{-3}$
\citep{bogdan2013}, which is comparable to that of the ICM in the
cluster outskirts, making their detection challenging. Worse, in cluster
cores the ICM density can be as high as $\sim$$10^{-2}\,\mathrm{cm}^{-3}$
\citep{zandanel2014}, which limits direct detection to only the most
luminous cluster galaxy halos \citep[e.g.,][]{sun2007}. While direct
detection likely is not feasible for the vast majority of a cluster's
galaxies, hot halos can be detected in a statistical sense by stacking
X-ray images of cluster members to derive the average X-ray emission
of a population of galaxies. This stacking method has been used to
detect diffuse hot halos around isolated field galaxies \citep[e.g.,][]{bogdan2015}.
In addition to the ICM, there are a number of other sources that contribute
to the overall X-ray emission around cluster galaxies. These include
low-mass X-ray binaries (LMXBs), high-mass X-ray binaries (HMXBs),
AGNs, and hot halos (if they are still around cluster galaxies).

In this work, we construct models that account for the dominant sources
of X-ray emission. This allows us to measure, in an average sense,
the X-ray luminosity of each component and enables us to infer whether
cluster members have retained their hot gas halos. To implement our
models, we stack archival \textit{Chandra X-ray Observatory} images
of the nearby \citep[$z=0.0622$;][]{vikhlinin2006} cluster Abell
1795 (A1795), and generate surface brightness profiles (SBPs) in two
clustercentric radius bins. A Markov chain Monte Carlo (MCMC) sampling
code is used to fit the SBPs and derive the contribution from each
model component.

Our data sets and sample selection are presented in Section\ \ref{sec:data},
while we describe our model and the results of our SBP fitting in
Section\ \ref{sec:models}. We discuss our results in Section\ \ref{sec:discussion},
and summarize our findings in Section\ \ref{sec:summary}. Throughout,
we adopt a WMAP7 cosmology \citep{komatsu2011} with ($\Omega_{\Lambda}$,
$\Omega_{M}$, $h$) = (0.728, 0.272, 0.704). In this cosmology, A1795
has a luminosity distance of 277.7 Mpc.

\section{Data and Sample Selection}

\label{sec:data}

\begin{figure}[!t]
\includegraphics{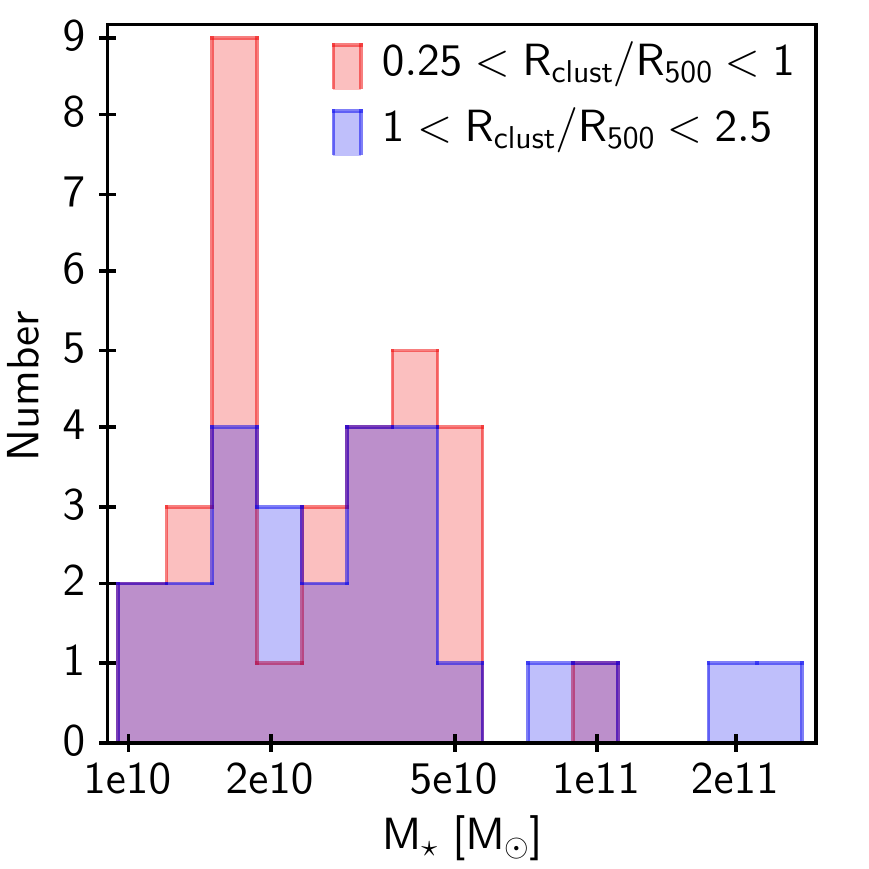}

\caption{Distribution of stellar masses for our final sample. The majority
of galaxies in both clustercentric radius bins are concentrated in
a fairly narrow range in stellar mass between $M_{\star}=10^{10}\,M_{\odot}$
(our lower mass cut) and $M_{\star}\sim5\times10^{10}\,M_{\odot}$.\label{fig:massDist}}
\end{figure}

\begin{figure}[!h]
\includegraphics[width=1\columnwidth]{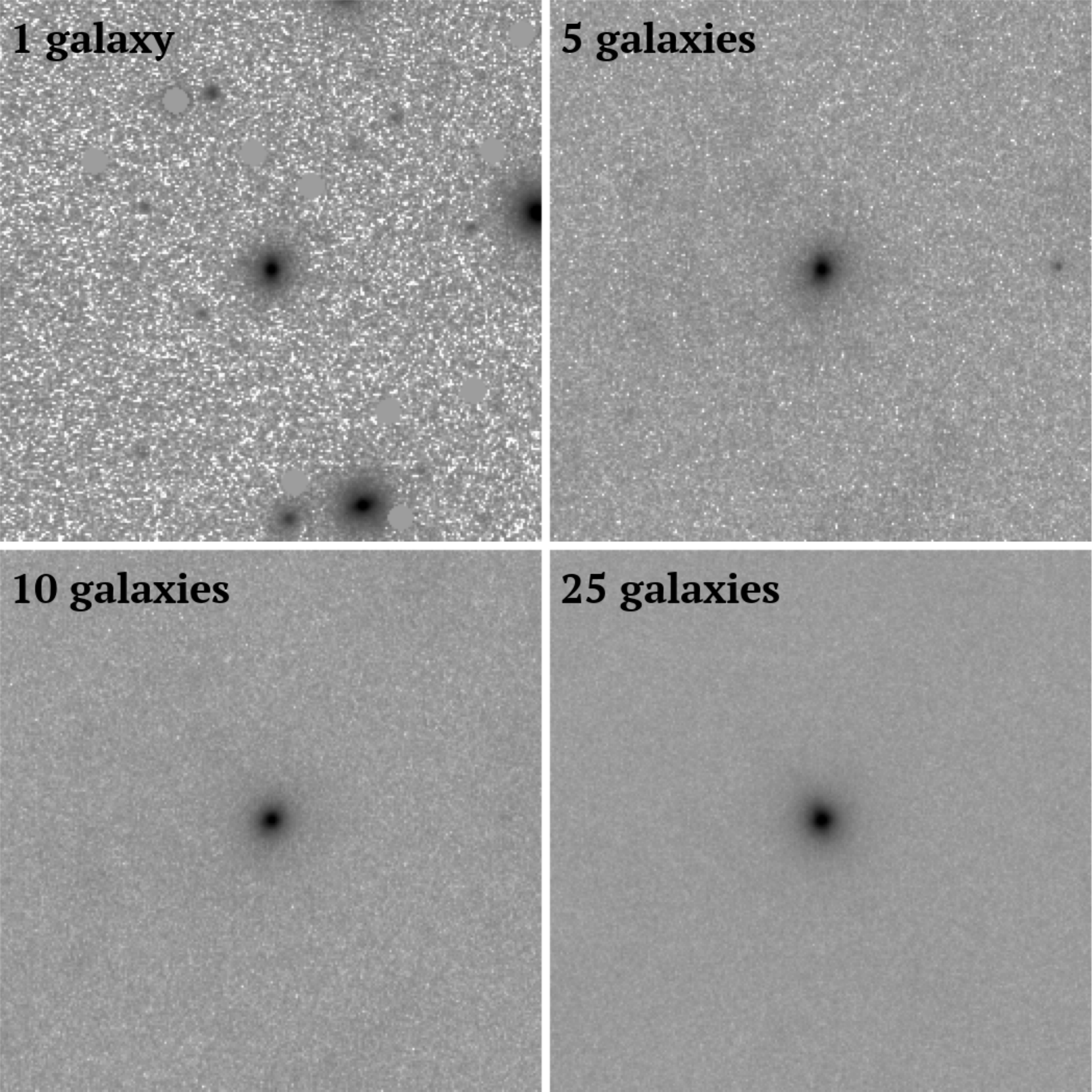}

\caption{Comparison of a single $i$-band image and median-stacked images with
five, 10, and 25 different galaxies. Each image is 260 pixels ($\sim$$103\arcsec$)
per side. Median stacking of the optical images results in a smooth
and relatively isotropic SBP. \label{fig:opticalStackComp}}
\end{figure}

\subsection{SDSS}

\label{subsec:SDSS_data}

The optical portion of our data comes from the Sloan Digital Sky Survey
(SDSS) Data Release 12 \citep{alam2015}. The SDSS provides positions
and spectroscopic redshifts for 864 galaxies located within $50\arcmin$
($\sim$3.6 Mpc) of A1795, whose position is taken from \citet{shan2015}.
Our cluster membership uses the redshift selection criterion from
\citet{shan2015}, which gives a redshift membership range of $0.0552<z_{\mathrm{gal}}<0.0692$
for A1795, based on its velocity dispersion. This cut, which results
in 190 galaxies, is very conservative and will likely exclude some
bonafide A1795 members. However, as background and foreground galaxies
will have hot halos, including them could bias our stacked X-ray images.

Using the $R_{500}$ for A1795 from \citet{vikhlinin2006}, members
are separated into two clustercentric radius ($R_{\mathrm{clust}}$)
bins: $0.25<R_{\mathrm{clust}}/R_{500}<1$ and $1<R_{\mathrm{clust}}/R_{500}<2.5$.
Galaxies within the inner 25\% of $R_{500}$ are excised to reduce
contamination from the ICM.

We match our member list to the catalog of \citet{chang2015}, who
measured star-formation rates (SFRs) and stellar masses using SDSS
and \textit{WISE} photometry for the spectroscopic SDSS sample compiled
by \citet{blanton2005}. This results in 135 members with SFRs and
stellar masses that are located within $0.25<R_{\mathrm{clust}}/R_{500}<2.5$.

Even though low-mass galaxies are the most numerous in galaxy clusters,
they will have their hot halos stripped more quickly than higher-mass
galaxies \citep{vijayaraghavan2015}. As our goal is to detect the
presence of hot halos around A1795 members, we impose a minimum stellar
mass cut of $10^{10}\,M_{\odot}$ on our sample. Figure\ \ref{fig:massDist}
shows the distribution of stellar masses for our final sample. The
remaining sample cuts will be described in Section\ \ref{subsec:chandra_data}.

SDSS $i$-band images that cover the A1795 field of view are downloaded,
then mosaicked using \textsc{Montage} \citep[version 5.0]{jacob2010b,jacob2010a}.\footnote{\url{http://montage.ipac.caltech.edu/}}
\textsc{SExtractor} \citep{bertin1996} is used to identify stars
in the mosaicked image and their positions are masked with zero values.
Individual $103\arcsec\times103\arcsec$ cutouts are made for each
cluster member from the $i$-band SDSS mosaic. For each clustercentric
radius bin, SDSS images are then stacked, and SBPs are calculated
(see Section\ \ref{subsec:SBP}). The $i$-band data will provide
us with a proxy for the low-mass X-ray binary populations within A1795
members (see Section\ \ref{subsec:XB_pop}).

We stack images by taking the median value at each pixel instead of
the mean. While the overall shape of the SBPs are similar (especially
towards the center of the stacks) between the two types of images,
those created by taking the median value at each pixel have smoother
shapes as they are not as impacted by outliers (i.e., light from non-target
galaxies). Figure\ \ref{fig:opticalStackComp} shows a comparison
of a single $i$-band image (top left) with median-stacked images
of multiple galaxies. While the image of the single galaxy is clearly
non-isotropic, individual galaxy features are smoothed over as more
galaxies are included in the stack. With 25 galaxies the stack is
highly smooth and circular.

We inspect the $i$-band images of the final A1795 members and find
10 of the 58 galaxies have obvious late-type features, with five in
each clustercentric radius bin. This results in early-type galaxy
fractions of $0.84_{-0.09}^{+0.07}$ and $0.81_{-0.11}^{+0.08}$ for
the inner and outer cluster, respectively.

\subsection{Chandra}

\label{subsec:chandra_data}

We have acquired 170 \chandra\ observations covering A1795 out to
$50\arcmin$ ($\sim$3.6 Mpc) from the \chandra\ Data Archive.\footnote{\url{http://cda.harvard.edu/chaser/}}
We match the observation positions against those of our galaxy sample
in order to select the 58 members that fall within $5\arcmin$ of
the \chandra\ optical axis in at least one observation (and also
meet the mass and clustercentric radius cuts described in the previous
section). As the \chandra\ point-spread function (PSF) is strongly
dependent upon the position within the telescope's field, degrading
as a function of off-axis angle, this selection helps mitigate PSF
broadening.

Our sample was observed on average four times per galaxy, for a total
sample exposure time of 5.1 Msec, with 3.4 Msec for the 32 galaxies
at low clustercentric radius and 1.7 Msec for the 26 galaxies in the
outer cluster.

\begin{figure}[!b]
\includegraphics[width=1\columnwidth]{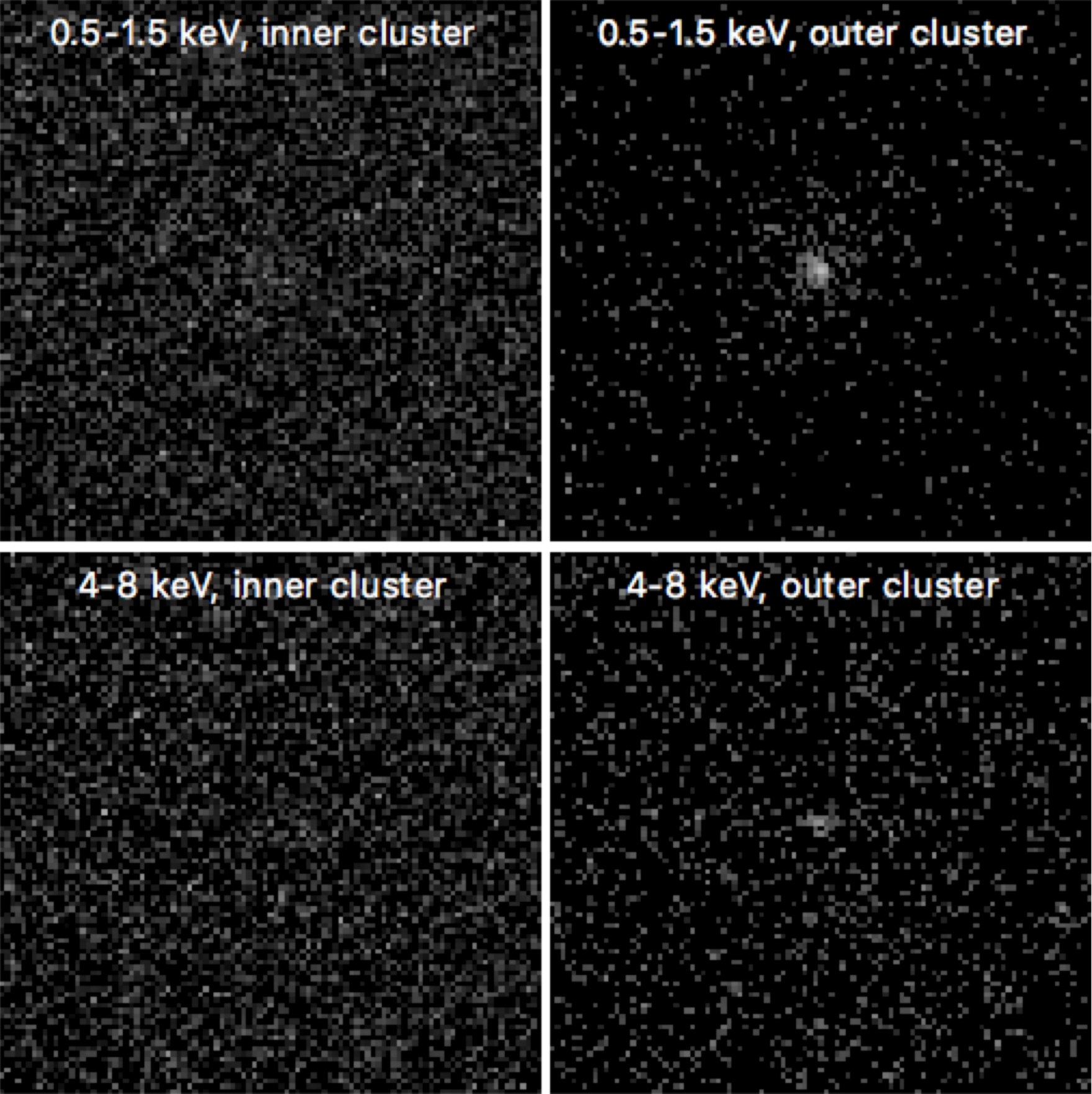}

\caption{Stacked X-ray images for the inner (left) and outer (right) cluster,
for both soft (top) and hard (bottom) X-rays. Each image is 101 pixels
($\sim$$50\arcsec$) per side. When compared to the outer cluster,
the inner cluster's relatively high background is evident at $0.5-1.5$
keV. Any potential gradient in the background due to telescope orientation
has removed through the stacking procedure.\label{fig:xrayStacks}}
\end{figure}

All observations are reprocessed using \textsc{CIAO} version 4.9 \citep{fruscione2006}
and version 4.7.6 of the \chandra\ Calibration Database. For each
observation, exposure maps are created with \textsc{CIAO}'s \texttt{fluximage}
script and PSF maps are generated with \texttt{mkpsfmap}. We then
run \texttt{wavdetect} to detect sources in each observation. For
each galaxy, we make $50\arcsec\times50\arcsec$ cutouts over the
0.5\textendash 1.5 keV (soft X-ray) and 4\textendash 8 keV (hard X-ray)
ranges from the exposure maps and the reprocessed events files. For
the galaxies that were observed more than once, cutouts are made from
each exposure map and events file in which the galaxy was observed.
All individual events file cutouts are then exposure corrected with
\texttt{fluximage}, using the script's default values, which ensures
that the exposure-corrected images have units of flux ($\mathrm{photons\ s^{-1}\ cm^{-2}}$).
Using \texttt{dmcopy} with the \texttt{exclude} filter, the source
positions detected above are masked in each exposure-corrected image.
We only exclude sources that lie more than $5\arcsec$ away from the
center of the cutout (i.e., the galaxy's location) to ensure that
we do not mask our final sample members. The exposure-corrected cutouts
are first stacked and averaged for each multiple-exposure galaxy,
then all cutouts are stacked and averaged for each clustercentric
radius bin. The final stacked images are shown in Figure \ref{fig:xrayStacks}.
We extract SBPs for each of these stacks, as described in the following
section. Because the exposure-corrected images have units of flux,
the SBPs are in units of flux per area.

In order to identify individually detectable objects we first estimate
our sensitivity in the soft X-ray images. For each cutout, we iteratively
add a single count to a random position in the image within a 5x5
pixel region ($\sim2.5\arcsec\times2.5\arcsec$). To ensure that we
are not adding flux to the target, counts are only added outside of
the central 50x50 pixels ($\sim$$25\arcsec\times25\arcsec$) of the
cutout. After each count has been added, we run \texttt{wavdetect}.
This process is repeated until a new source has been detected, at
which point we exposure correct the modified cutout. The background
flux of the cutout is measured using a large area that does not intersect
with the target, artificial source, or any sources originally detected
in the image. The region defined by \texttt{wavdetect} is used to
sum the flux of the added source, which is converted into a luminosity
with $L_{X}=4\pi d_{\mathrm{L}}^{2}f_{X}$, where $d_{\mathrm{L}}$
is the luminosity distance of A1795 and $f_{\mathrm{X}}$ is the X-ray
flux.

Background-subtracted luminosities are calculated for sources detected
in each cutout, as long as the source is located at the center of
the image. A target is flagged as detected if its background-subtracted
luminosity is at least as large as the sensitivity limit found above.
Four such objects are found, all lying beyond $R_{500}$. Formally,
our detection fractions are $0.00_{-0}^{+0.06}$ and $0.15_{-0.07}^{+0.11}$
for the inner and outer cluster, respectively.

We investigate fitting the SBPs of the individually detected galaxies.
For three of the galaxies, there are not enough data to generate SBPs
with fine enough binning to accurately model the inner portions of
the profile. The remaining galaxy has substantially more data, but
upon fitting the galaxy's SBP, the residuals are so large that it
prevents any meaningful analysis. However, we find no evidence in
any of the four profiles of an extended component beyond what could
be expected from a combination of an AGN, LMXBs, and the ICM. While
these galaxies likely contribute a significant portion of the data
in the stacked profile, we cannot analyze them individually. Hence,
we do not remove them from our stacked outer cluster SBP; this is
akin to measuring SFRs with stacked infrared observations.

\subsection{Measuring Surface Brightness Profiles}

\label{subsec:SBP}

We define 52 (15) concentric annuli for the stacked \chandra\ (SDSS)
images, centered on the middle of the pixel that corresponds to the
target's location. In general, the SBPs are then defined by summing
the flux in all pixels in each annulus, then dividing by the area
of the annulus. Starting with the innermost annulus and working outward,
pixels are considered to be in an annulus if the center of the pixel
is contained within the annulus' radius. The entire flux from that
pixel is then considered to be in that annulus. The sum of the flux
from all the pixels in each annulus is divided by the area, which
is the total number of pixels. This initial surface brightness is
then divided by the square of the pixel scale to give a surface brightness
in terms of square arcseconds. For fitting and plotting purposes,
the corresponding radius point for each surface brightness value is
redefined as the midpoint between the current and previous annulus
radius.

\section{Modeling the X-ray Emission From Member Galaxies in Abell 1795}

\label{sec:models}

In this section we will describe some of the components that can be
expected to contribute to X-ray emission in and around galaxies in
A1795, with the initial simplifying assumption that A1795 members
no longer possess a hot halo. Hence, we expect X-ray emission only
due to AGNs (Section\ \ref{subsec:AGN_PSF}), the ICM (Section\ \ref{subsec:ICM}),
and/or X-ray binaries (XRBs; Section\ \ref{subsec:XB_pop}). To test
our assumption, we will build models that incorporate these components
and assess how well they fit the X-ray SBPs of stacked A1795 members.
We shall refer to the position within the galaxies' stacked images
as the galactocentric radius, $R_{\mathrm{gal}}$.

\subsection{Active Galactic Nuclei}

\label{subsec:AGN_PSF}

In our SBPs, the emission due to an AGN takes the shape of the \chandra\
PSF. To accurately model the PSF at various positions on the detector,
we use the \textsc{SAOTrace} \citep[v2.0.4;][]{jerius1995}\footnote{\url{http://cxc.harvard.edu/cal/Hrma/SAOTrace.html}}
ray tracing code to simulate light rays through the telescope optics,
and \textsc{MARX} \citep[v5.3.2;][]{davis2012}\footnote{\url{http://space.mit.edu/cxc/marx/}}
to create PSF images. The resulting PSF images\textemdash one for
each galaxy for each observation in which it was observed\textemdash are
multiplied by the total flux in a $1\arcsec$-radius aperture around
the target position in the respective reprocessed observations. For
each galaxy, all of its PSF images are stacked and averaged. As with
the X-ray images, these resulting images are also stacked and averaged
after being separated into the two clustercentric radius bins. SBPs
as a function of galactocentric radius, $I_{\mathrm{PSF}}\left(R_{\mathrm{gal}}\right)$,
are then extracted for each of the PSF stacks, following the method
described in Section\ \ref{subsec:SBP}.

In order to account for PSF differences between optical and X-ray
data, we generate a convolution kernel, $PSF$, from each PSF SBP,
which will be used to convolve some components in our models.

\subsection{The Intracluster Medium}

\label{subsec:ICM}

A constant component is used in all of our X-ray fits to account for
the cluster background. The background in individual X-ray cutouts
may have a gradient that depends on the galaxy's position in the cluster
and the orientation of the telescope at the time of observation. However,
our average stacked images of multiple galaxies over multiple observations
effectively removes any such gradient, as can be seen by the relatively
isotropic background in all four panels of Figure \ref{fig:xrayStacks}.

\subsection{X-ray Binaries}

\label{subsec:XB_pop}

To determine the expected relative significance of each type of XRB
system in A1795 galaxies we use relations from the literature to estimate
their X-ray luminosities. In each case, we scale the relation from
the energy range in the published work to our energy range of interest,
0.5\textendash 1.5 keV, using WebPIMMS.\footnote{\url{https://heasarc.gsfc.nasa.gov/cgi-bin/Tools/w3pimms/w3pimms.pl}}
The HMXB spectrum is modeled as a power law with a photon index of
2.1 \citep{sazonov2017}, while the LMXB spectrum is treated as a
7 keV thermal Bremsstrahlung \citep{boroson2011}.

For HMXBs, we use the relation between SFR and HMXB X-ray luminosity
from \citet[their Eq.$\,39$]{mineo2012}, which, when scaled to our
energy range, gives an expected HMXB luminosity of
\begin{equation}
L_{0.5-1.5\mathrm{\,keV}}^{\mathrm{HMXB}}/\left(\mathrm{erg\,s^{-1}}\right)=1.06\times10^{39}\,SFR/\left(M_{\odot}\,\mathrm{yr}^{-1}\right).\label{eq:expectedHMXB}
\end{equation}
To find the expected X-ray luminosity due to LMXBs, we use the relation
between stellar mass and LMXB X-ray luminosity from \citet{zhang2012}.\footnote{$L_{0.5-8\,\mathrm{keV}}/M_{\star}=9.6\times10^{39}\,\mathrm{erg/s}$
per $10^{11}\,M_{\odot}$} Scaled to our desired energy range, this results in an expected LMXB
luminosity of 
\begin{eqnarray}
L_{0.5-1.5\,\mathrm{keV}}^{LMXB}/\left(\mathrm{erg\,s^{-1}}\right) & = & 2.48\times10^{39}\,M_{\star}/\left(10^{11}M_{\odot}\right).\label{eq:expectedLMXB}
\end{eqnarray}

\begin{figure}[!t]
\includegraphics{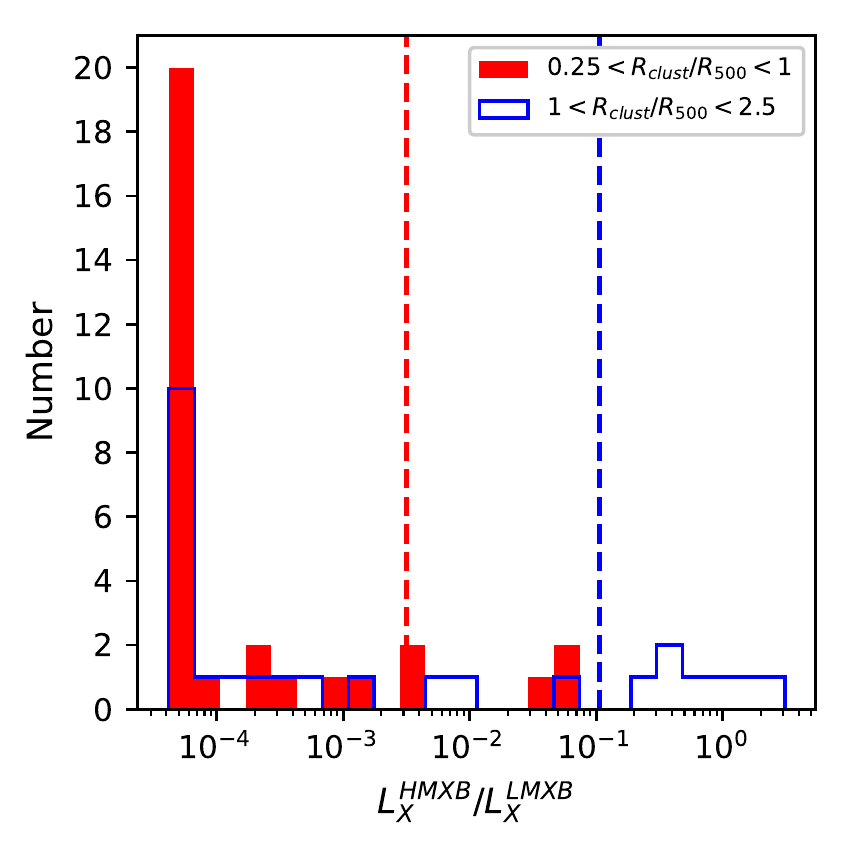}

\caption{Distribution of the ratio between the expected 0.5\textendash 1.5
keV luminosity of HMXBs and LMXBs. The mean ratio of 0.003 (0.1) for
the inner (outer) cluster is shown by the red (blue) vertical dashed
line. This figure demonstrates that LMXBs are expected to dominate
the stellar contribution to the X-ray profiles, and we can safely
exclude HMXBs from our analysis.\label{fig:xrbDist}}
\end{figure}

\begin{figure}[!t]
\includegraphics{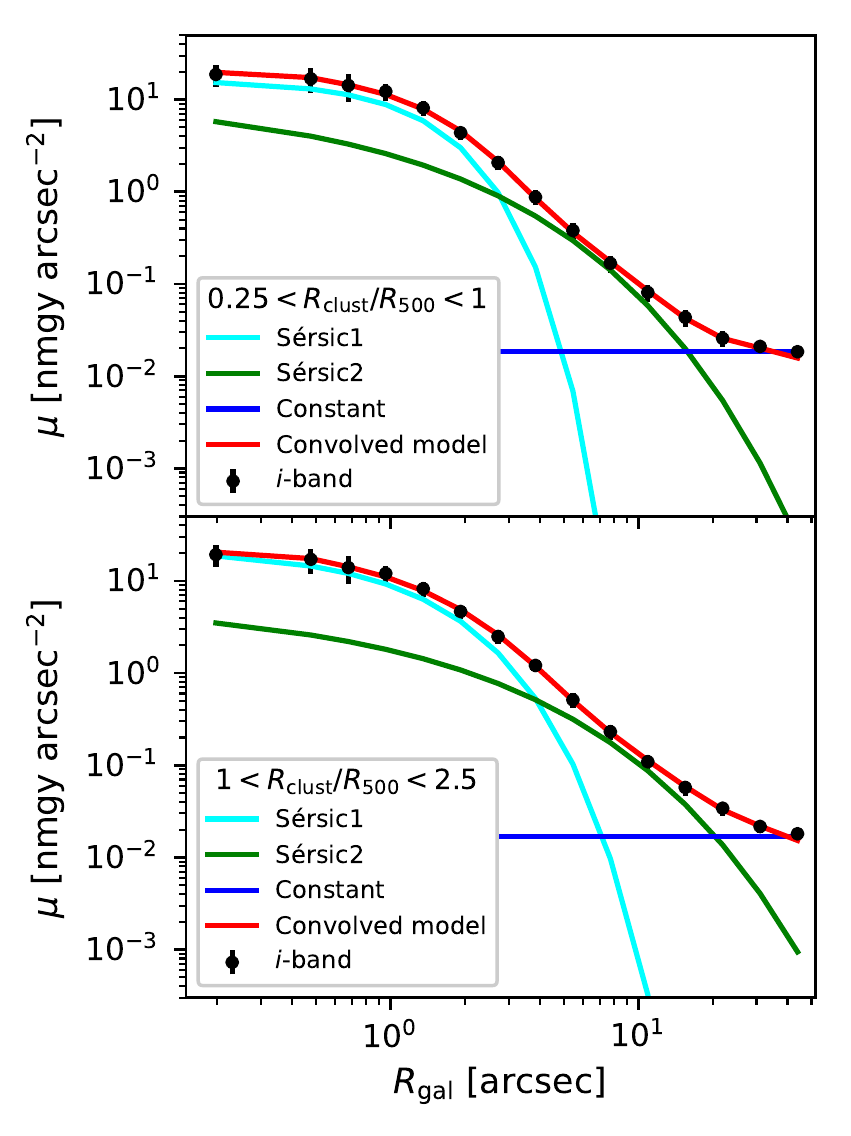}

\caption{Optical surface brightness of stacked A1795 members (black points
and error bars) in the $i$-band. SBPs in each panel are well fit
with a model comprising two S\'{e}rsic functions and a constant.
The resulting two-S\'{e}rsic component provides the shape of the
LMXB component of our model.\label{fig:sb_opt}}
\end{figure}

We calculate the ratio of the expected 0.5\textendash 1.5 keV luminosity
due to HMXBs and LMXBs ($L_{0.5-1.5\mathrm{\,keV}}^{\mathrm{HMXB}}/L_{0.5-1.5\,\mathrm{keV}}^{LMXB}$)
for our final sample and plot their distribution in Figure \ref{fig:xrbDist}.
The mean ratio in each clustercentric radius bin is shown by the dashed
vertical line. For outer cluster galaxies, the expected luminosity
due to HMXBs is an order of magnitude lower than that of LMXBs. In
the inner cluster, HMXBs are expected to contribute less than one
percent of the XRB flux. Because of their substantially lower expected
0.5\textendash 1.5 keV luminosity, we can safely exclude HMXBs from
our analysis.

The companion in an LMXB system is typically an older, low-mass star.
Since cluster galaxies are comprised mainly of old stellar populations,
and LMXBs are not expected to be distributed differently than the
underlying old stellar population, we will use the distribution of
old stars in A1795 galaxies to model the shape of the LMXB contribution.
The $i$-band SDSS images provide excellent proxies for the old stellar
populations.

We perform a least squares fit to each of the SDSS SBPs measured in
Section\ \ref{subsec:SDSS_data}, modeling the profiles with two
S\'{e}rsic \citep{sersic1963} functions plus a constant. During
fitting, the sum of the S\'{e}rsic functions and the constant is
convolved with a Gaussian whose standard deviation is approximately
the median value of the SDSS $i$-band PSF ($1.4\arcsec$).\footnote{\url{http://www.sdss.org/dr12/imaging/other_info/}}
The results of these fits are shown in Figure \ref{fig:sb_opt}. The
non-convolved double S\'{e}rsic function, $S_{i}\left(R_{\mathrm{gal}}\right)$,
is used to describe the shape of the LMXB component in our X-ray emission
model. The LMXB ($i$-band) surface brightness component is
\begin{equation}
I_{i}\left(R_{\mathrm{gal}}\right)=\left[S_{i}\left(R_{\mathrm{gal}}\right)\ast PSF\right].\label{eq:SBP_LMXB}
\end{equation}
In this equation, the S\'{e}rsic function is convolved with the PSF
kernel.

\subsection{Modeling X-ray Emission With an AGN, LMXB, and Background\label{sec:baselineModel}}

\begin{figure*}[t]
\includegraphics{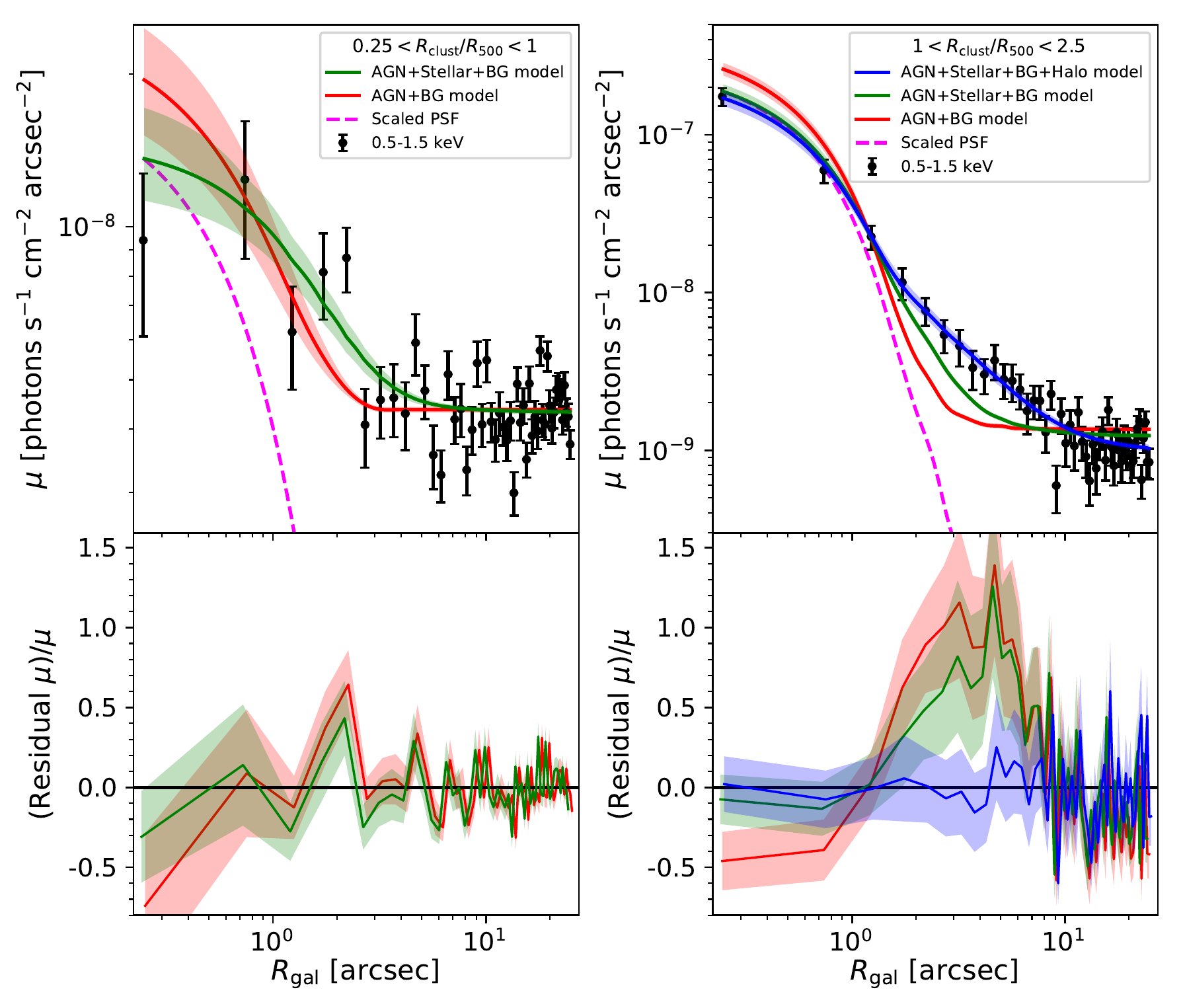}

\caption{Upper panels: SBPs of stacked 0.5\textendash 1.5 keV images of A1795
members (black points and error bars). Red curves are the median SBPs
of the AGN+BG model, which comprises a background (constant), and
an AGN (simulated \chandra\ PSF). The green curves show the SBPs
of the AGN+Stellar+BG model, which include an additional LMXB ($i$-band
SDSS) component. The blue curve in the upper right panel is the median
SBP of the AGN+Stellar+BG+Halo model, which further adds a hot halo
component (Beta function). The shaded regions show the 15th to 85th
percentile range of the model SBPs. All model SBPs are generated through
MCMC simulations. For both outer cluster runs that have an LMXB component
(green and blue curves), the distribution of $i$-band-to-X-ray ratios
from the inner cluster \textsc{MCMC} run with the AGN+Stellar+BG model
(green curve) is used as a prior. The dashed magenta lines show the
simulated PSF, scaled so that the innermost point equals the innermost
point of the AGN+Stellar+BG model. This shows that an AGN-only model
is not sufficient to account for the X-ray profile in either clustercentric
radius bin. Lower panels: Relative residuals ({[}data-model{]}/model)
of the measured and model SBPs. Relative residuals are color coded
to match their respective model, the uncertainty ranges are shaded,
and the lines are slightly shifted to the left and right for clarity.
\label{fig:mcmcComp_split}}
\end{figure*}

Before describing our models we first present the 0.5\textendash 1.5
keV SBPs (black points and error bars) in the upper panels of Figure
\ref{fig:mcmcComp_split}. These SBPs are measured from the stacked
images shown in Figure\ \ref{fig:xrayStacks}, using the procedure
outlined in Section\ \ref{subsec:SBP}. While centralized emission
can be seen in the stacked inner cluster SBP, it is not immediately
obvious in the upper left panel of Figure\ \ref{fig:xrayStacks}.
Although we do not present it, a smoothed image makes the emission
apparent.

The magenta curves in Figure\ \ref{fig:mcmcComp_split} show the
ray-traced PSFs that we modeled in Section\ \ref{subsec:AGN_PSF}.
Given that the shapes of the PSFs (and the AGN emission they imply)
are fixed, and all that can be adjusted are their amplitudes, it is
clear that an AGN alone cannot account for all the flux in and around
A1795 members, in either clustercentric radius bin.

The arbitrarily scaled PSF (AGN) reasonably fits (by eye) the first
few SBP values at low $R_{\mathrm{gal}}$ for both the inner and outer
cluster galaxies. Qualitatively, the main deficit in an AGN-only model
is that it cannot account for the background flux (the relatively
flat SBP values at large $R_{\mathrm{gal}}$). To remedy this we define
our first model,
\begin{equation}
I\left(R_{\mathrm{gal}}\right)=C+A_{\mathrm{AGN}}I_{\mathrm{PSF}}\left(R_{\mathrm{gal}}\right),\label{eq:AGB+BG}
\end{equation}
where $C$ is the (constant) ICM background, and $A_{\mathrm{AGN}}$
is the multiplicative factor for $I_{\mathrm{AGN}}\left(R_{\mathrm{gal}}\right)$,
the surface brightness contribution due to an AGN. With this AGN+BG
model defined, we use MCMC sampling to determine the contribution
of each component, enabling us to generate model SBPs. We use the
MCMC sampling code \textsc{emcee} \citep{foreman-mackey2013},\footnote{\url{http://dan.iel.fm/emcee/current/}}
which uses sets of Markov chain walkers to explore the parameter space,
with each walker starting from some initial assigned value. To determine
initial values of $A_{\mathrm{AGN}}$ and $C$, we take the amplitude
of our scaled PSF and the median of the 20 outermost SBP values, respectively.
We use 400 walkers per parameter, each starting at a random value
within 0.1\% of the initial input. Using uniform priors on our parameters
($A_{\mathrm{AGN}}\geq0$ and $C\geq0$), we run \textsc{emcee} for
1600 steps and cut the first 100 steps for burn-in. This results in
$6\times10^{5}$ sets of parameters (i.e., $6\times10^{5}$ values
for each parameter). This process is the same for both the inner and
outer cluster.

We generate an SBP for each parameter set and find the 15th, 50th
and 85th percentiles of these profiles at each galactocentric radius.
The sampled SBPs are plotted in red in the upper panels of Figure\ \ref{fig:mcmcComp_split}.
The lower panels show the relative residuals between the AGN+BG model
and 0.5\textendash 1.5 keV SBP ({[}data - model{]}/model). We use
the Python package \textsc{uncertainties}\footnote{\url{https://pythonhosted.org/uncertainties/}}
to derive errors on the relative residuals by propagating the uncertainties
on the measured and model SBPs. For the model SBP, we treat the 15th
and 85th percentiles as the uncertainty range. The relative residual
is plotted with a solid line and the uncertainty range is shaded.

Qualitatively, the AGN+BG model fits the background fairly well in
both the inner and outer cluster, but the overall fit is poor. More
specifically, the surface brightness is overestimated at low $R_{\mathrm{gal}}$,
while being underestimated at moderate $R_{\mathrm{gal}}$ (around
$\sim$$2\arcsec$ in the inner cluster and $2\arcsec\lesssim R_{\mathrm{gal}}\lesssim7\arcsec$
in the outer cluster). This is due to the shape of the model interior
to the background being driven only by the PSF. Using the median model
SBPs, we calculate reduced $\chi^{2}$ values of $3.02$ and $4.26$
for the inner and outer cluster, respectively. These values suggest
what was also determined qualitatively: an AGN+BG model is not sufficient
to account for the X-ray flux of A1795 members.

We now consider a model that includes all the components previously
described in this section: the cluster background, LMXBs, and AGNs.
We reiterate our initial assumption that none of the X-ray emission
is due to a hot gas halo from individual galaxies. The general form
of this AGN+Stellar+BG model is

\begin{equation}
I\left(R_{\mathrm{gal}}\right)=C+A_{i}I_{i}\left(R_{\mathrm{gal}}\right)+A_{\mathrm{AGN}}I_{\mathrm{PSF}}\left(R_{\mathrm{gal}}\right),\label{eq:AGN+Stellar+BG}
\end{equation}
where $I_{i}\left(R_{\mathrm{gal}}\right)$ is the $i$-band SBP from
Equation\ \ref{eq:SBP_LMXB}, $A_{i}$ is the multiplicative factor
modulating the $i$-band S\'{e}rsic, and $A_{\mathrm{AGN}}$ and
$C$ are defined as in Equation\ \ref{eq:AGB+BG}.

With our AGN+Stellar+BG model defined, we again use \textsc{emcee},
beginning with the inner cluster. To determine the initial parameter
values, we first perform a least squares fit between the 0.5\textendash 1.5
keV SBP and the AGN+Stellar+BG model. For this fit, and during the
inner cluster MCMC simulation, the only constraints on the parameters
are that they must be non-negative. The resulting three best-fit parameters
of Equation \ref{eq:AGN+Stellar+BG} for the inner cluster are used
as initial values for \textsc{emcee}. For this model, we use 500 walkers
per parameter, assigning their starting values as before. \textsc{emcee}
is run for 2500 steps, cutting the first 500 for burn-in, resulting
in one million sets of parameters.

The upper left panel of Figure \ref{fig:mcmcComp_split} presents
the sampled inner cluster SBP for the AGN+Stellar+BG model in green.
The relative residuals for this model are plotted with the same color
in the bottom left panel. Qualitatively, the overall fit is good due
to the extra flexibility provided by the LMXB term. The large residuals
that were present in the AGN+BG model have been reduced slightly.

Quantitatively, the $\chi^{2}$ drops from 147.7 for the AGN+BG model
to 138.2 for the AGN+Stellar+BG model. This drop of 9.5 in $\chi^{2}$
with only one less degree of freedom (through the addition of the
LMXB's $A_{i}$ term) results in a subtle improvement in the reduced
$\chi^{2}$, which decreases from 3.02 to 2.88 for the inner cluster
AGN+Stellar+BG model. The qualitatively better fit of the AGN+Stellar+BG
model, coupled with its lower reduced $\chi^{2}$, validates the addition
of the LMXB term. While a reduced $\chi^{2}$ of 2.88 may suggest
that an additional component could be added to the model, it might
also be a symptom of the scatter of the background. We will revisit
the topic of potentially adding more terms to this model for the inner
cluster, but we must first turn to the outer cluster.

To determine initial parameter values for the MCMC sampling at $R_{\mathrm{clust}}>R_{500}$,
we perform a least squares fit between the AGN+Stellar+BG model and
the SBP of the outer cluster. Since $A_{i}$ is effectively a scaling
factor between the $i$-band and X-ray emission, it should be independent
of position within the cluster, so during this fit we fix $A_{i}$
at the value found during the least squares fit of the inner cluster.
This results in best-fit parameters for the background and AGN components,
which we assign as the initial values for the \textsc{emcee} simulation.
We use 500 walkers for each of the three parameter in the outer cluster
\textsc{emcee} run, assigning them for the background and AGN components
as we did for the inner cluster. For the amplitude of the LMXB component,
we draw 500 random values from the inner cluster $A_{i}$ distribution.
In addition to ensuring that the amplitude of the AGN and constant
components are non-negative, we also use the inner cluster $A_{i}$
distribution as a prior and require $A_{i}$ to fall between the minimum
and maximum of the inner cluster distribution. Following our procedure
for the inner cluster, we run \textsc{emcee} for 2500 steps for the
outer cluster, cutting the first 500 steps for burn-in. Using the
resulting one million sets of parameters we generate one million AGN+Stellar+BG
model SBPs. The sampled model SBPs (relative residuals) are plotted
in green in the upper (lower) right panel of Figure\ \ref{fig:mcmcComp_split}.

The AGN+Stellar+BG model fits the data well at low galactocentric
radius, with relative residuals consistent with zero for $R_{\mathrm{gal}}\lesssim1\arcsec$.
However, the residuals increase to order unity as the model cannot
accurately account for the X-ray flux between $\sim$$2\arcsec$ and
$\sim$$10\arcsec$. With a reduced $\chi^{2}$ of 2.60, the AGN+Stellar+BG
model better fits the measured SBP than just the AGN+BG. However,
it clearly cannot account for all of the 0.5\textendash 1.5 keV flux.

Allowing $A_{i}$ to freely vary between clustercentric radius bins
may produce a better fit in the outer cluster. However, this would
imply that the scaling relation between the $i-$band and the X-ray
is a function of clustercentric radius, and we can invoke no physical
basis for such a claim. While the reduced $\chi^{2}$ of 2.60 may
be in part due to the scatter of the data, the discrepancy between
$\sim$$2\arcsec$ and $\sim$$10\arcsec$ suggests a shortcoming
of the AGN+Stellar+BG model. 

Qualitatively, it is clear that the AGN+Stellar+BG model does not
accurately fit the SBP in the outer cluster. This result, along with
a reduced $\chi^{2}$ of 2.60, suggests that augmenting this model
with an additional component is warranted in order to more adequately
account for all the X-ray emission from $R_{\mathrm{clust}}>R_{500}$
A1795 galaxies.

To test whether the order in which we perform the fits (inner cluster
followed by outer cluster) is the cause of the low quality fit of
the AGN+Stellar+BG model in the outer cluster, we repeat the analysis
with this model, but perform the fits and MCMC simulations on the
outer cluster before the inner cluster. We do not present these results
but note that the AGN+Stellar+BG model fits the SBP in the outer cluster
slightly better than with the original analysis order. However, there
is still an excess of flux near $10\arcsec$, and the fit of the SBP
in the inner cluster is much worse, more so than with the AGN+BG model.
Hence, we conclude that the order of the fit is not the cause of the
large flux excess in the AGN+Stellar+BG fit of the outer cluster.

\begin{figure*}[!t]
\includegraphics{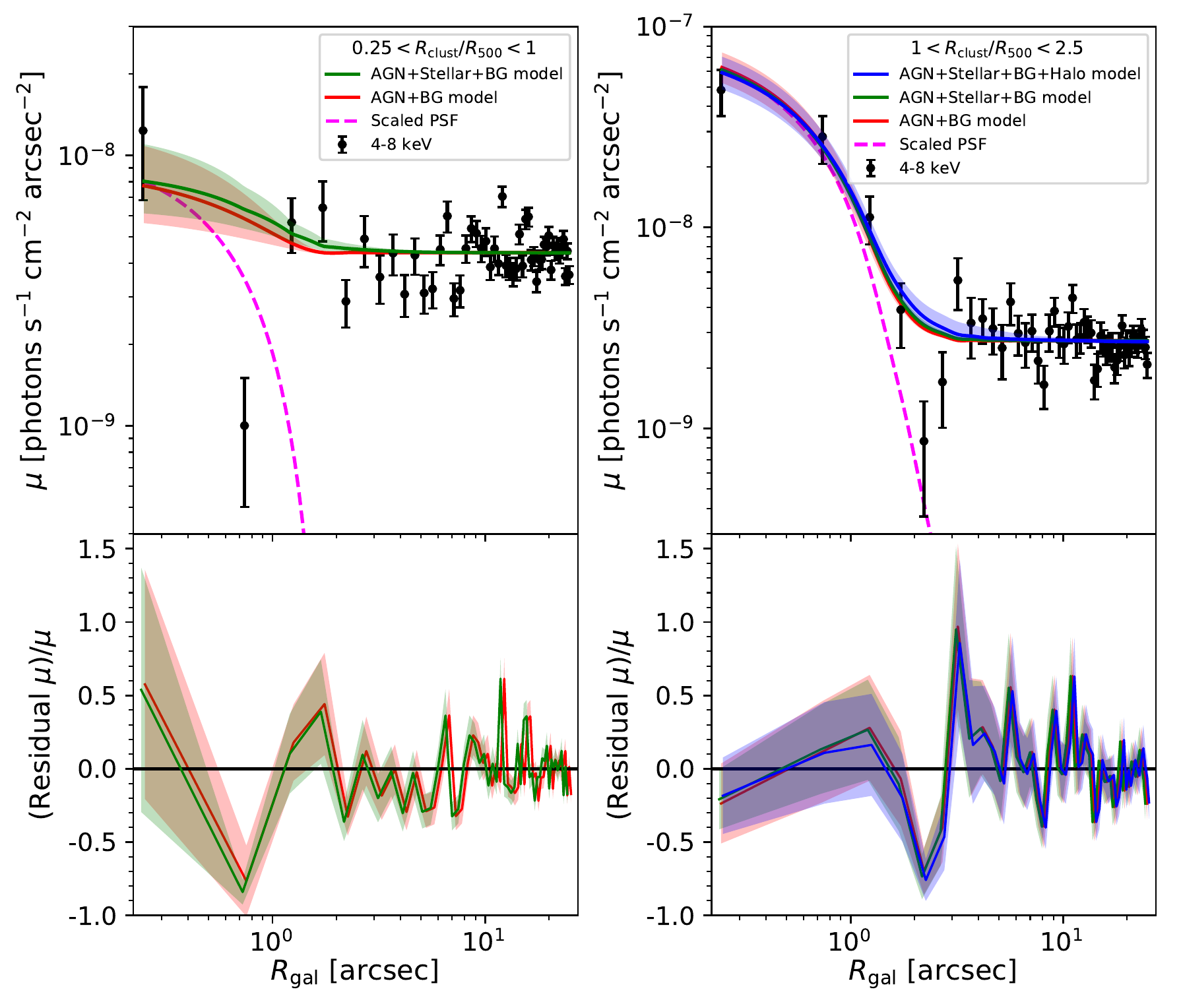} 

\caption{Same as Figure~\ref{fig:mcmcComp_split} but for 4\textendash 8 keV
data. All three models fit the SBP in both clustercentric radius bins
equivalently well. This suggests that the AGN+Stellar+BG (AGN+Stellar+BG+Halo)
model is an appropriate model of the hot halo emission in the soft
X-rays in the inner (outer) cluster. Unlike at lower energy, the AGN+BG
models fit the hard X-ray data extremely well at all clustercentric
radii, suggesting that an AGN and the cluster background are the primary
sources of 4\textendash 8 keV emission. \label{fig:mcmcComp_split_hardXray}}
\end{figure*}

\subsection{Modeling the X-ray Excess}

\label{subsec:SBP_beta}

In order to better account for the X-ray flux excess at $2\arcsec\lesssim R_{\mathrm{gal}}\lesssim10\arcsec$
in the outer cluster, we now modify our AGN+Stellar+BG model by including
a Beta ($\beta$) model \citep{cavaliere1978}, which has a surface
brightness profile of 
\begin{equation}
I\left(R_{\mathrm{gal}}\right)=I_{0}\left[1+\left(\frac{R_{\mathrm{gal}}}{R_{c}}\right)^{2}\right]^{-3\beta+1/2}\label{eq:betaFun}
\end{equation}
\citep{anderson2014}, where $I_{0}$ is the central surface brightness
(when $R=0$) and $R_{c}$ is the core radius. The $\beta$ model
is commonly used to model hot halos around galaxies \citep{anderson2014},
and its inclusion is motivated in part by the results of \citet{vijayaraghavan2015},
who found, using synthetic X-ray observations, that hot halos should
be detectable in stacked low-energy (0.1\textendash 1.2 keV) galaxy
images out to $\sim$$10\arcsec$ at $z=0.05$. Given the location
of the flux excess in our outer cluster SBPs, the $\beta$ model seems
like the ideal choice to augment the AGN+Stellar+BG model.

As with the S\'{e}rsic component in the AGN+Stellar+BG model, the
$\beta$ component is convolved with the PSF surface brightness kernel.
In our new model, which we will call the AGN+Stellar+BG+Halo model,
$I_{0}$, $R_{c}$, and $\beta$ are free parameters, with the constraints
that $I_{0}$ must be non-negative, $\beta$ can vary between 0.3
and 0.9, and the core radius must be in the range $0.08\arcsec<R_{c}<8.5\arcsec$.

Following the procedure outlined in Section \ref{sec:baselineModel},
we find the best fit between the X-ray surface brightness in the outer
cluster and the AGN+Stellar+BG+Halo model, using the best-fit parameter
values as inputs into \textsc{emcee}. As with the AGN+Stellar+BG model
\textsc{emcee} run, we use the $A_{i}$ distribution of the inner
cluster as a prior, also drawing from it to seed the initial walker
values.

Given that we have three additional parameters for this model, we
use 2500 walkers and run \textsc{emcee} for $1\times10^{5}$ steps.
The amount of runtime required for the walkers to fully explore the
parameter space is much larger than with any of the previous models,
and we discard all but the last 5000 steps of the run for burn-in.
This results in 12.5 million sets of parameters that are used to generate
SBPs, which are plotted in blue in the upper right panel of Figure\ \ref{fig:mcmcComp_split},
with the relative residuals shown in the bottom right panel, with
the same color.

Qualitatively, the fit in the outer cluster with this new model is
substantially better than with either of the previous two models.
All of the X-ray excess at $2\arcsec\lesssim R_{\mathrm{gal}}\lesssim10\arcsec$
has been accounted for. With a reduced $\chi^{2}$ of 1.62, the fit
is also quantitatively superior and validates the addition of the
beta component in the outer cluster.

Given these positive results, we repeat the analysis on the inner
cluster with the AGN+Stellar+BG+Halo model. While we do not plot the
modeled SBP, qualitatively the fit is consistent with the results
of the AGN+Stellar+BG model. Furthermore, with a $\chi^{2}$ of 139.1,
the quality of the fit is no better than with the AGN+Stellar+BG model,
and with the addition of the three Beta parameters, the reduced $\chi^{2}$
increases to 3.09. These results suggest that a hot halo component
is unnecessary for fitting the 0.5\textendash 1.5 keV SBPs of inner
cluster A1795 members.

\subsection{Hard X-ray SBPs}

In this section, we repeat our analysis for hard X-ray (4\textendash 8
keV) SBPs of A1795 members to test whether our soft X-ray model components
are appropriately modeling their intended physical counterparts. Specifically,
we focus on the hot halo in the outer cluster. Given the temperature
of galaxy hot halos \citep[$\sim10^7$ K;][]{forman1985}, which corresponds
to $kT\sim0.9$ keV, their expected X-ray emission should fall primarily
within our soft X-ray window. If modeled correctly, we should expect
that the Beta component of our outer cluster AGN+Stellar+BG+Halo model
would account for little-to-no hard X-ray flux.

In Section\ \ref{subsec:XB_pop}, we use a fixed 7 keV thermal Bremsstrahlung
model to represent the spectrum of LMXBs. However their emission is
expected to span a broad range in energy \citep[e.g., 0.3--8 keV;][]{boroson2011}
and could potentially impact the fit in the hard X-ray band.

Following the procedure described in Sections\ \ref{sec:baselineModel}
and \ref{subsec:SBP_beta}, we analyze the 4\textendash 8 keV SBPs
of A1795 members. In the upper panels of Figure\ \ref{fig:mcmcComp_split_hardXray}
we plot the measured hard X-ray SBPs, the sampled model SBPs, and
the scaled PSFs. The relative residuals of the models are plotted
in the lower panels. In this figure, we use the same color scheme
as we did for the soft X-ray analysis.

As with the soft X-ray SBPs, the scaled PSF alone cannot account for
the 4\textendash 8 keV flux. Qualitatively, however, the AGN+BG model
fits the hard X-ray SBPs in both the inner and outer cluster, despite
the large amount of scatter present in the data. In both clustercentric
radius bins, the AGN+Stellar+BG model is consistent with the AGN+BG
model. In the inner cluster, the addition of the LMXB component results
in an increase in the reduced $\chi^{2}$ from 5.09 for the AGN+BG
model to 5.63 for the AGN+Stellar+BG model. In the outer cluster,
all three models fit the 4\textendash 8 keV SBPs equivalently well.
Because of this, each additional component in the model worsens the
reduced $\chi^{2}$. The AGN+BG model has a reduced $\chi^{2}$ of
1.75. With the addition of the LMXB component, the reduced $\chi^{2}$
rises to 1.84. Finally, the AGN+Stellar+BG+Halo model has the worst
reduced $\chi^{2}$ of the three outer cluster models, with a value
of 2.08.

This can be interpreted as the Halo component adding no new information
to the model fits at 4\textendash 8 keV. Hence we are confident that
the Beta function is an appropriate choice for modeling any potential
hot halo contribution to the 0.5\textendash 1.5 keV SBPs of stacked
outer cluster A1795 members.

\subsection{Soft X-ray Luminosities of Model Components}

\begin{figure}[!t]
\includegraphics{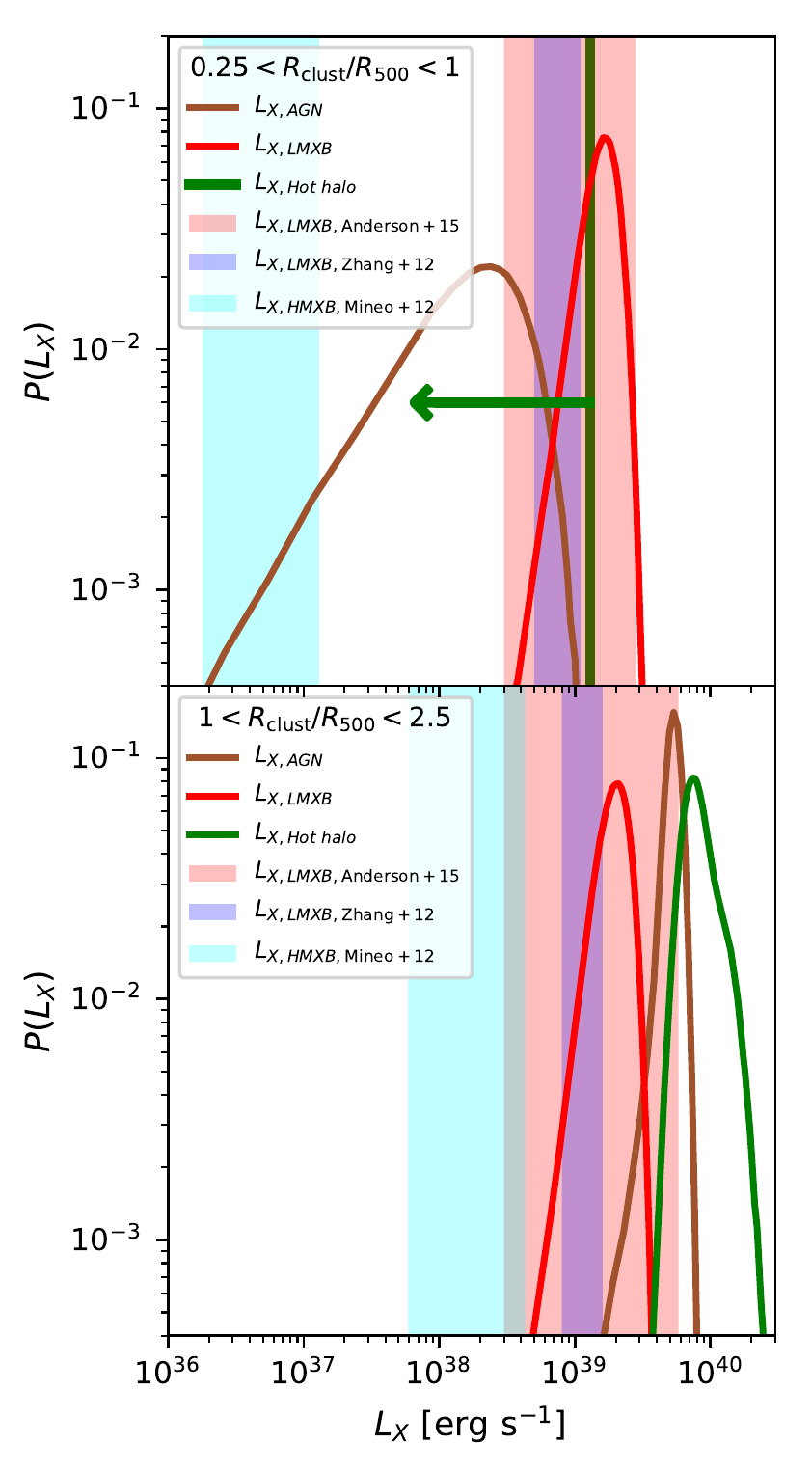}

\caption{X-ray luminosity probability distributions of non-background components
of the AGN+Stellar+BG model (inner cluster; upper panel) and the AGN+Stellar+BG
model (outer cluster; lower panel) The green vertical line in the
upper panel shows the estimated upper limit on hot halo luminosity.
X-ray luminosities are calculated from the parameter distributions
that result from the MCMC simulations. The blue shaded regions show
the expected $L_{X}$ due to LMXBs based on the stellar mass-$L_{X,LMXB}$
relation from \citet{zhang2012}. The red shaded region shows the
expected $L_{X}$ due to LMXBs from \citet{anderson2015}, which they
based on the $L_{X,LMXB}$-$L_{K}$ relation from \citet{boroson2011}.
The LMXB X-ray luminosity derived through our MCMC simulations (red
histograms) are consistent with both comparison LMXB X-ray luminosity
ranges. The cyan shaded region shows the expected $L_{X}$ due to
HMXBs from the SFR-$L_{X,HMXB}$ relation from \citet{mineo2012}.
All X-ray luminosities presented here are in the 0.5\textendash 1.5
keV range. \label{fig:LxDist_1betaFree}}
\end{figure}

While the AGN+Stellar+BG+Halo model provides a substantially better
fit of the 0.5\textendash 1.5 keV SBP in the outer cluster, the quality
of the fit alone does not tell us much about whether A1795 galaxies
retain their hot halos.

In order to determine the luminosity of the hot halo (and other components),
we use the entire post-burn in set of parameters from our MCMC simulations
to generate SBPs for each component in both models (AGN+Stellar+BG
for the inner cluster and AGN+Stellar+BG+Halo for the outer cluster).
The profiles are integrated to find the total flux contributed by
each component and the luminosity of each component is calculated.
The probability distributions of the non-background components are
shown in Figure \ref{fig:LxDist_1betaFree}. There are clear detections
of all components in both clustercentric radius bins.

The LMXB components have similar probability distributions in the
low- and high-clustercentric radius bins, which is unsurprising, given
that we use the inner cluster distribution of $A_{i}$ as a prior
for the MCMC run in the outer cluster. This is justified as the scaling
between the X-ray and $i$-band should not vary based on location
within the cluster.

Given our data and possible model components, we consider the AGN+Stellar+BG
model to be the ideal representation of the soft X-ray SBP in the
inner cluster. For completeness, however, we estimate the upper limit
on any potential hot halo luminosity by taking the 90th percentile
of the set of post-burn in $\beta$ function parameters from the test
run of the AGN+Stellar+BG+Halo model in the inner cluster (see the
last paragraph in Section\ \ref{subsec:SBP_beta}). This value, $L_{X,\mathrm{halo}}=1.3\times10^{39}$
erg s$^{-1}$, is plotted as the green vertical line in the upper
panel of Figure\ \ref{fig:LxDist_1betaFree}. While the probability
distribution of the hot halo component in the outer cluster has a
peak near $10^{40}$ erg s$^{-1}$, it does have a non-negligible
probability tail that extends to a few $10^{39}$ erg s$^{-1}$. However,
the minimum value in the outer cluster hot halo luminosity distribution
is $L_{X,\mathrm{halo}}=1.7\times10^{39}$ erg s$^{-1}$, approximately
1.3 times larger than the upper limit of the inner cluster hot halo
luminosity. Based on these two data points, there appears to be an
environmental trend in the X-ray luminosity of A1795 galaxy hot halos.

In Table\ \ref{tab:softLuminosity}, we present the non-background
component luminosities for each model, the background-subtracted total
luminosity (Total), and the upper limit to the hot halo luminosity
for the inner cluster.

\begin{table}
\caption{0.5\textendash 1.5 keV Luminosities\label{tab:softLuminosity}}

\centering{}%
\begin{tabular}{lc}
\addlinespace[-0.02\textheight]
 & \tabularnewline
\midrule
\midrule 
Component & $L_{X}$\tabularnewline
 & ($10^{39}\,\mathrm{erg\,s^{-1}}$)\tabularnewline
\midrule
\multicolumn{2}{c}{Inner Cluster}\tabularnewline
Total & $1.8\pm1.7$\tabularnewline
LMXB & $1.6\pm0.5$\tabularnewline
AGN & $0.2_{-0.1}^{+0.2}$\tabularnewline
Hot halo & $<1.3$\tabularnewline
\midrule 
\multicolumn{2}{c}{Outer Cluster}\tabularnewline
Total & $15.6\pm2.6$\tabularnewline
LMXB & $2.0\pm0.5$\tabularnewline
AGN & $5.4_{-0.8}^{+0.7}$\tabularnewline
Hot halo & $8.1_{-3.5}^{+5}$\tabularnewline
\bottomrule
\end{tabular}
\end{table}

\section{Discussion}

\label{sec:discussion}

\subsection{Literature Comparison}

\subsubsection{Total and Component X-ray Luminosity}

\label{subsec:modelComponents}

As noted previously, the LMXB components of our models have similar
luminosities in the inner and outer cluster. While the LMXB component
at $0.25<R_{\mathrm{clust}}/R_{500}<1$ provides the dominant contribution
to the total background-subtracted X-ray luminosity, LMXBs are the
subdominant component in the outer cluster. In order to determine
whether our values are reasonable, we use the stellar masses of our
final sample and Equation\ \ref{eq:expectedLMXB} to find the expected
range in 0.5\textendash 1.5 keV luminosity for the LMXBs in the inner
and outer cluster (shown in Figure\ \ref{fig:LxDist_1betaFree} with
the blue shaded regions). Furthermore, \citet{anderson2015} provides
expected LMXB luminosities based on the $L_{X,LMXB}$-$L_{K}$ relation
from \citet{boroson2011}. Converting their published $L_{X}$ values
to the 0.5\textendash 1.5 keV band, we plot these with the red shaded
regions in Figure\ \ref{fig:LxDist_1betaFree} for the stellar mass
range of our final sample. This is remarkable consistency between
three different estimates of LMXB luminosity.

The cyan shaded regions in Figure\ \ref{fig:LxDist_1betaFree} show
the expected 0.5\textendash 1.5 keV luminosity due to HMXBs from Equation\ \ref{eq:expectedHMXB},
based on the SFRs of our final sample. This further exemplifies that
HMXBs are not expected to provide a substantial contribution to the
SBPs of A1795 members, and supports their exclusion from our models.

\citet{anderson2015} provide total galaxy X-ray luminosities (0.5\textendash 2.0
keV) for ``locally brightest'' SDSS galaxies, binned by stellar
mass. In order to compare these to our background-subtracted $L_{X}$
values, we find the mean stellar mass of our inner and outer cluster
samples, with uncertainties on the mass derived through 10000 iterations
of bootstrap resampling with replacement. In the inner (outer) cluster,
the log of the mean stellar mass is $\log\left(M_{\star}/M_{\odot}\right)=10.48\pm0.05$
($\log\left(M_{\star}/M_{\odot}\right)=10.67_{-0.11}^{+0.09}$). For
our inner cluster galaxies, the luminosity of the corresponding mass
bins from \citet{anderson2015} spans the range $7.2\times10^{38}<L_{X,\mathrm{Total}}/\mathrm{erg\,s^{-1}}<3.1\times10^{39}$,
while the range for the bins that overlap with our outer cluster stellar
mass is $3.1\times10^{39}<L_{X,\mathrm{Total}}/\mathrm{erg\,s^{-1}}<9.9\times10^{40}$.
These luminosities are corrected for the difference in energy ranges
using WebPIMMS, assuming a compound model of a 1 keV thermal Bremsstrahlung
for the hot gas, 7 keV thermal Bremsstrahlung for LMXBs, and a power
law with a photon index of 2 for the AGN and HMXB emission. We do
not consider the uncertainties published by \citet{anderson2015},
although these would only serve to widen the luminosity ranges.

\citet{kim2013} measured 0.3\textendash 8 keV luminosities of nearby
gas-poor early-type galaxies. Following the same procedures as with
the \citet{anderson2015} luminosities, we convert the \citet{kim2013}
values to our soft X-ray energy band, and find that they span the
range $1.1\times10^{39}<L_{X,\mathrm{Total}}/\text{\ensuremath{\mathrm{erg\,s^{-1}}}}<5.4\times10^{39}$.
This range is consistent with our inner cluster background-subtracted
luminosity of $L_{X,\mathrm{Total}}=\left(1.8\pm1.7\right)\times10^{39}\,\mathrm{erg\,s^{-1}}$.
Given that these galaxies have at most negligible hot halo emission
(on average), this consistency is to be expected. Outer cluster A1795
members, on the other hand, have a relatively strong hot halo component
so we would not expect their total luminosity ($L_{X,\mathrm{Total}}\sim2\times10^{40}\,\mathrm{erg\,s^{-1}}$)
to agree with those of gas-poor galaxies. A simple subtraction of
the hot halo luminosity from the total for outer cluster galaxies
gives a value that is also consistent with the results of \citet{kim2013}.

The overall consistency between these total $L_{X}$ values from the
literature and our background-subtracted total X-ray luminosities
is encouraging as it suggests that our selection of a constant for
the ICM was the correct one.

The X-ray luminosity of outer cluster AGNs ($L_{\mathrm{X,\mathrm{AGN}}}\sim5\times10^{39}\,\mathrm{erg\,s^{-1}}$)
is consistent with the range of AGN luminosities found by \citet{lamassa2012}
for Seyfert 2 galaxies. While our AGN luminosity at $1<R_{\mathrm{clust}}/R_{500}<2.5$
falls on the low end of this range ($2.5\times10^{39}\lesssim L_{\mathrm{X,\mathrm{AGN}}}/\mathrm{erg\,s^{-1}}\lesssim1.6\times10^{42}$;
converted to the 0.5\textendash 1.5 keV band), it is likely that only
a few of the 26 galaxies in this bin have strong AGN emission as the
fraction of AGNs in low-redshift clusters is small; \citet{martini2009}
and \citet{haines2012} both found X-ray detected AGN fractions of
$<1\%$ in $0.05<z<0.3$ and $0.15<z<0.3$ clusters, respectively.
Even at high clustercentric radius few cluster galaxies are expected
to host AGNs (e.g., \citet{lopes2017} found an AGN fraction of $\sim$5\%
for $R/R_{200}>1$ galaxies in $z<0.1$ clusters).

While AGNs provide a substantial component of the X-ray luminosity
at large clustercentric radius, they are clearly subdominant for low
clustercentric radius galaxies ($L_{\mathrm{X,\mathrm{AGN}}}\sim2\times10^{38}\,\mathrm{erg\,s^{-1}}$).
Their X-ray luminosity is an order of magnitude lower than that of
LMXBs, and even lower than the upper limit of hot halo luminosity.
The large difference in AGN $L_{X}$ as a function of clustercentric
radius is unsurprising as the AGN fraction tends to increase with
clustercentric radius \citep[e.g.,][]{ehlert2014,lopes2017}.

\subsubsection{Hot Halos Around Abell 1795 Galaxies}

\begin{figure*}[!t]
\includegraphics{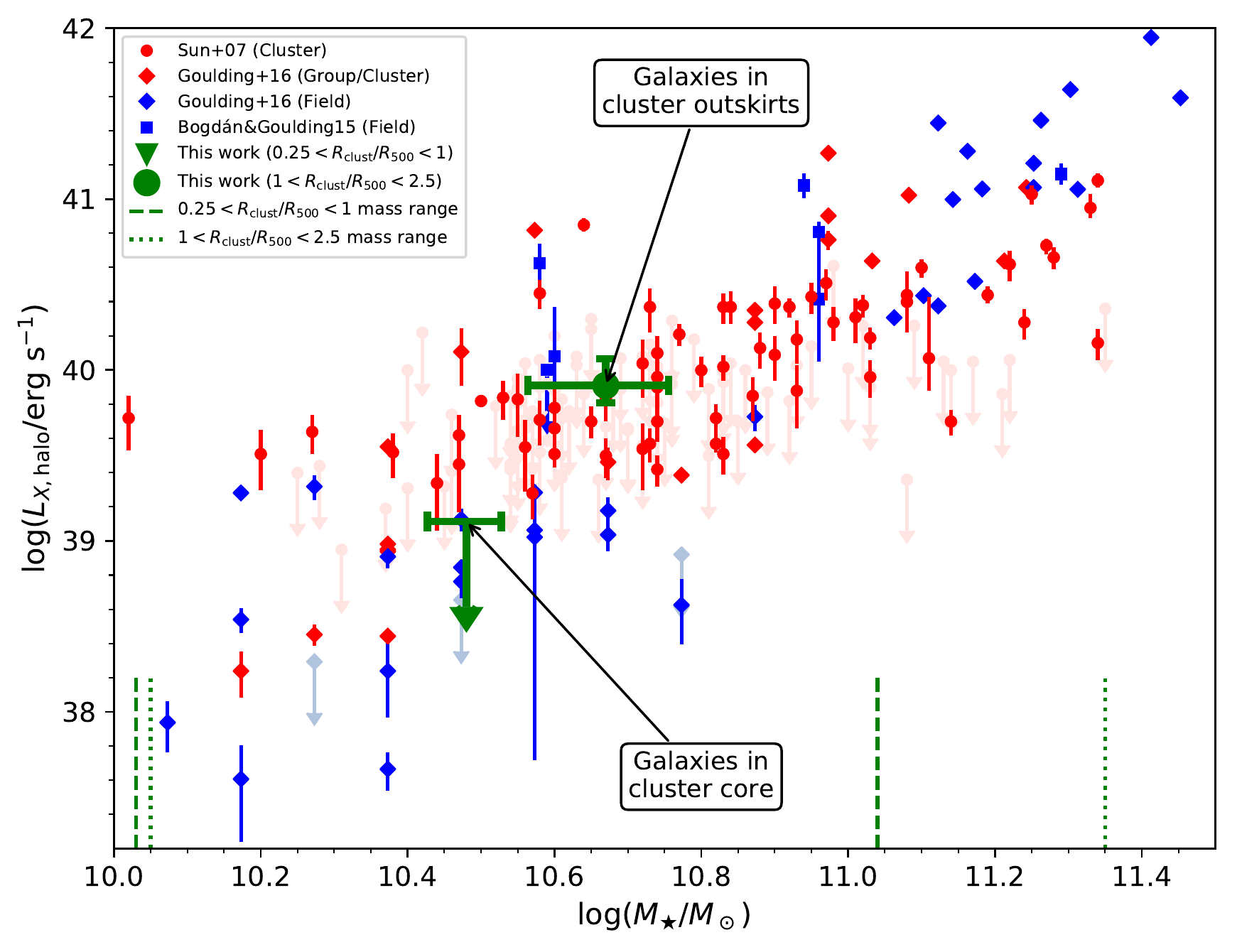}

\caption{Soft X-ray (0.5\textendash 1.5 keV) hot halo luminosity of versus
galaxy stellar mass. The green point (vertical error bar) shows the
median (15th to 85th percentile) X-ray luminosity of the hot halo
of outer cluster members of A1795. The downward facing green arrow
represents the estimated upper limit of the hot halo of inner cluster
A1795 members. Both of our values are plotted at the mean stellar
mass of the galaxies in the stack, and the horizontal error bars are
the standard error on the mean stellar mass of the stacked galaxies,
derived through 10000 iterations of bootstrap resampling. The short
vertical lines at the bottom of the plot show the mass range of A1795
members. Other points represent a selection of hot halo $L_{X}$ values
from the literature, scaled to the 0.5\textendash 1.5 keV band. The
red points are for cluster galaxies, while the blue points are for
galaxies in the field. Lighter colored red and blue points with downward
facing arrows represent upper limits.\label{fig:Lx_Mstar}}
\end{figure*}

In Figure\ \ref{fig:Lx_Mstar} we plot the 0.5\textendash 1.5 keV
luminosity (upper limit) of the hot halo component of our AGN+Stellar+BG+Halo
(AGN+Stellar+BG) model at the mean stellar mass found in Section\ \ref{subsec:modelComponents}.
We compare our values to a selection of hot halo luminosities from
the literature, for field and cluster galaxies (see legend). Here
we briefly describe the comparison samples. The reader is encouraged
to refer to the cited papers for further details. For all comparison
samples, we convert their published $L_{K}$ values to stellar masses
using the $K$-band stellar mass-to-light ratio from \citet{bell2003},
$\log_{10}\left(M/L_{K}\right)=-0.42+0.033\log_{10}\left(M_{\star}h^{2}/M_{\odot}\right)$,
and all galaxies with $\log\left(M_{\star}/M_{\odot}\right)<10$ are
removed from our comparison samples. We also convert all luminosities
from their published energy range to our soft (0.5\textendash 1.5
keV) X-ray band.

We plot the luminosity of cluster galaxy halos from \citet{sun2007}
with the red circles. Most of the hot halo luminosities from \citet{sun2007}
are for individual detections, although in some cases multiple (two
to four) fainter galaxies in the same cluster were stacked if they
had similar net counts and resided close to each other on the \chandra\
detector. \citet{sun2007} provided the cluster in which their galaxies
reside as well as the angular distance between the galaxies and center
of the cluster's ICM for the majority of their sample. We found $R_{500}$
for 24 of the 25 clusters they studied and calculated $R/R_{500}$
for the 156 galaxies in these clusters that had a published angular
distance. All of these galaxies lie within $\sim$0.9 Mpc of the ICM
center, corresponding to our low clustercentric radius bin. For consistency
with our sample, we remove \citet{sun2007} galaxies with $R/R_{500}<0.25$.

\citet{bogdan2015} hot halo $L_{X}$ values are plotted with blue
squares. Their sample comprises 3130 field elliptical galaxies at
$0.01<z<0.05$, separated into different mass and velocity dispersion
bins, with the X-ray images in each bin stacked.

We include hot halo luminosities from \citet{goulding2016} for galaxies
from the MASSIVE survey \citep{ma2014}. For these galaxies \citet{goulding2016}
include the number of nearby neighbors for each. We separate their
galaxies into two samples: those that have at least five neighbors
(their classification for rich groups and clusters) are considered
as group/cluster galaxies, and galaxies with fewer than five neighbors
are considered field galaxies.\footnote{Increasing the cutoff to 15 neighbors had no impact on the sample.}
We remove group/cluster galaxies that they classify as either being
the central in their potentials or as hosting an AGN. We also include
a subset of lower-mass ATLAS\textsuperscript{3D} \citep{cappellari2011}
galaxies that \citet{goulding2016} published as a comparison sample.
We select galaxies that were also studied by \citet{su2015} and use
their classifications to separate them: galaxies that are listed as
a Virgo cluster member, have at least 15 nearby SDSS neighbors, or
were defined as a group/cluster galaxy in the literature are considered
cluster galaxies; all others are considered field galaxies. The comparison
cluster/group (field) galaxies from \citet{goulding2016} are shown
with red (blue) diamonds.

At $0.25<R_{\mathrm{clust}}/R_{500}<1$, A1795 galaxies have a stacked
$L_{X,\mathrm{halo}}$ upper limit almost uniformly lower than all
other cluster galaxies plotted. Clearly, hot halos with low X-ray
luminosity do exist in the cluster environment, but there appears
to be a lack of cluster galaxies with $\log\left(M_{\star}/M_{\odot}\right)\lesssim10.8$
that host hot halos with low field-relative $L_{X,\mathrm{halo}}$.
This could be due to observational biases: cluster galaxies are embedded
in the ICM, making individual detections of very low X-ray luminosity
hot halos nearly impossible. Observations of field galaxies are not
subject to this high background, making hot halo detections easier.
Another possible cause of the paucity of low-luminosity hot halos
around lower-mass cluster galaxies is that they are unable to hold
onto their tenuous halo gas when traveling through the ICM.

While we do not have a formal detection of $L_{X,\mathrm{halo}}$
for the inner cluster, by stacking 32 galaxies with 3.4 Msec of \chandra\
coverage, we are able to make measurements not typically possibly
for individual cluster galaxies. Our low clustercentric radius hot
halos have a stacked X-ray luminosity upper limit that is consistent
with the lowest values plotted for field galaxies.

If the true inner cluster $L_{X,\mathrm{halo}}$ value is somewhat
near its upper limit, there are two broad possibilities for the distribution
of individual hot halo luminosities: most (or all) of the galaxies
in the inner cluster could have some small residual amount of hot
halo gas; or a few members may possess a relatively large amount of
hot halo gas. However, due to the small number of galaxies in our
sample, and the coarseness of our binning, determining which, if any,
of these two scenarios is beyond the scope of this work.

In the outer cluster, A1795 galaxies have hot halo luminosities consistent
with the comparison samples around the same stellar mass. These galaxies
have almost uniformly higher $L_{X,\mathrm{halo}}$ than field and
cluster galaxies with $10<\log\left(M_{\star}/M_{\odot}\right)\lesssim10.5$,
and towards the upper envelope in luminosity for galaxies at $10.5\lesssim\log\left(M_{\star}/M_{\odot}\right)\lesssim11.3$.
Given that our hot halo luminosity is the average for 26 galaxies
over both of these mass ranges, these results suggests that A1795
galaxies beyond the $R_{500}$ have large hot gas halos.

While massive cluster galaxies appear to hold onto their hot halos
more readily than their less massive cluster counterparts, the highest
mass ($\log\left(M_{\star}/M_{\odot}\right)\gtrsim11$.1) field galaxies
have almost uniformly higher hot halo X-ray luminosities than the
most massive cluster galaxies. However, since field galaxies are not
subject to the ram pressure of an ICM, it is unsurprising that they
can build up larger hot gas reserves. Hence, these differences in
hot halo X-ray luminosity of field and cluster galaxies at large stellar
mass are likely due to environmental effects, while at lower stellar
mass the exact cause may be a combination of environments effects
and observation biases.

\subsection{Implications on the Quenching of Cluster Galaxies}

We have shown, based on their hot halo X-ray luminosities, that A1795
members are a dichotomous population. On average, galaxies within
$R_{500}$ retain little-to-none of their hot halo gas. Outer cluster
galaxies have substantially more luminous hot halos, with X-ray luminosities
a factor of six larger than the upper limit of the inner cluster $L_{X,\mathrm{halo}}$
(see Table\ \ref{tab:softLuminosity} and Figure\ \ref{fig:Lx_Mstar}).
With only two very large clustercentric radius bins, we cannot determine
the radial trend with any accuracy, however it is likely that we are
observing the removal of hot halos as galaxies fall into A1795. Taken
on its own, this is evidence for ongoing strangulation.

Given these results, a next step would be to investigate the implications
on different quenching mechanisms. While an in depth study of stellar
populations and cold gas content of A1795 members is beyond the scope
of this work and would require more extensive photometry, we can further
leverage the \citet{chang2015} catalog. It provides specific SFRs
($sSFR=SFR/M_{\star}$), which are commonly used to determine whether
a galaxy is quiescent or star forming. Adopting an sSFR cutoff of
$\log\left(sSFR/\mathrm{Gyr^{-1}}\right)>-1$ \citep[e.g.,][]{lin2014},
we find that none of our galaxies are classified as star-forming.
If we instead use a less conservative cut of $\log\left(sSFR/\mathrm{Gyr^{-1}}\right)\gtrsim-1.5$
\citep[see their Figure$\ $A1 for $10\lesssim\log(M_{\star}/M_{\odot})\lesssim10.5$ galaxies at $z\sim0.1$]{genel2018},
two of our outer cluster galaxies are classified as star forming;
none of the inner cluster galaxies, however, make this more relaxed
cut. This results in quiescent fractions of $1.0_{-0.06}^{+0}$ (32/32)
and $0.92_{-0.09}^{+0.05}$ (24/26) for the inner and outer cluster,
respectively. With only $\sim$30 galaxies in each bin, the fractions
are quite uncertain and formally consistent with each other, so we
are unable to measure any radial trend.

Using the $R_{200}$ for A1795 from \citet{shan2015}, we can convert
our $R/R_{500}$ to $R/R_{200}$ values and compare our quiescent
fractions to those of \citet{wetzel2012}, who measured quiescent
fraction as a function of $R_{200}$ for $z\sim0.045$ galaxies in
$\log\left(M_{\mathrm{halo}}/M_{\odot}\right)>14$ halos (their Figure\ 5).
In terms of $R_{200}$, our inner cluster galaxies span $0.15\lesssim R/R_{200}\lesssim0.61$.
Over this range, our fraction is comparable to the values plotted
by \citet{wetzel2012}, being consistent with two of their $\sim$6
binned fractions. At large clustercentric radius ($0.61\lesssim R/R_{200}\lesssim1.51$),
the situation is similar, with our fraction consistent with one of
their $\sim$4 binned values.

A galaxy that no longer possesses a hot halo yet is still actively
star-forming would be strong evidence for ongoing strangulation. However,
the nature of this work precludes such a discovery. While we can identify
at most two galaxies in the outer cluster that are still forming stars,
given our method for measuring X-ray luminosities of model components,
we cannot determine the \textsl{individual} strengths of these galaxies'
hot halos. Even though most, if not all, A1795 members are quiescent,
we can still begin to broadly investigate some of the possible quenching
mechanisms at play.

Since none of the inner cluster members are still forming stars, this
is actually evidence against \textit{ongoing} strangulation. However,
it is possible that members had their hot halos stripped, but retained
a portion of their cold gas when they entered the cluster environment.
That cold gas could have been subsequently consumed as the galaxies
made their way to the inner cluster. This scenario is supported by
the results of \citet{zinger2018}, who found that RPS is not an effective
mechanism for removing gas from galactic disks in cluster outskirts.

The ineffectiveness of RPS at large clustercentric radius would also
suggest that ram pressure is likely not the cause of the quiescence
of the outer cluster members. Additionally, the galaxies in the outskirts
of A1795 still have, on average, substantial hot halos, so ram pressure
may not yet be strongly affecting them. However, inner cluster members
have negligible hot halos, which may indicate that RPS has effectively
removed the majority of the hot halo gas \textit{and} cold interstellar
gas. Since ram pressure is proportional to the square of a galaxy's
velocity, the effects of RPS will likely be stronger near the centers
of clusters, where galaxies are traveling more quickly through the
ICM.

Given the high X-ray luminosity of the AGN component in our model
at large clustercentric radius, quenching due to an AGN may be a possibility.
As we noted in Section\ \ref{subsec:modelComponents}, though, only
a handful of cluster galaxies, even in the outskirts, are likely to
host an AGN, so AGN quenching on a large scale in A1795 is unlikely.
With its large mass ($M_{500}=5.46\times10^{14}\,M_{\odot}$), A1795
would not be a conducive environment for galaxy-galaxy interactions,
which tend to favor regions with low galaxy velocities. Some A1795
members at large clustercentric radius have likely been recently accreted
from lower density (group) environments, where tidal interactions
between galaxies are more common. It would seem that regardless of
the dominant quenching mechanism, a substantial fraction of the quiescent
galaxies, particularly at large clustercentric radius, may have arrived
in the cluster pre-quenched (pre-processing).

\section{Summary}

\label{sec:summary}

In this paper, we model the stacked, soft (0.5\textendash 1.5 keV)
X-ray emission of spectroscopic member galaxies in A1795. We model
the extended, stacked emission using a combination of spatially-flat
background cluster emission, central emission from an AGN, extended
emission from a stellar component (i.e., LMXBs), and an additional
extended component corresponding to the diffuse, hot halo associated
with individual galaxies.

As an ensemble, galaxies interior to $R_{500}$ have total (i.e.,
background-subtracted) soft X-ray luminosities that are consistent
with nearby gas-poor early-type galaxies. In contrast, galaxies exterior
to $R_{500}$ are 3\textendash 14 times brighter in the X-ray than
that of the comparison sample. With a 0.5\textendash 1.5 keV luminosity
of $L_{X,\mathrm{halo}}=\left(8.1_{-3.5}^{+5}\right)\times10^{39}\,\mathrm{erg\,s^{-1}}$,
extended hot halos have been detected around A1795 members exterior
to $R_{500}$, in a statistical sense. This hot halo luminosity also
accounts for the difference in total luminosities between the outer
cluster members and the comparison gas-poor early-type galaxies. While
hot halos provide a significant component of the X-ray emission of
outer cluster members, we find that they are clearly subdominant in
the inner cluster, where we calculate an upper limit of $L_{X,\mathrm{halo}}<1.3\times10^{39}\,\mathrm{erg\,s^{-1}}$.

Such a large difference in the X-ray luminosity of extended gas halos
around member galaxies interior and exterior to $R_{500}$ suggests
that we are witnessing the stripping of hot halos from A1795 members
as they travel through the dense ICM. On its own, this result would
support quenching by on-going strangulation. However, all of the inner
cluster members are already quiescent according to their sSFRs, so
at most we can suggest that they were quenched by strangulation. While
outer cluster members, on average, still possess their hot halos,
nearly all are quiescent. This quenching was likely caused before
the galaxies entered the cluster environment, with the removal of
the hot halo preventing the ``reignition'' of star formation in
the future. Pre-processing is the preferred quenching explanation
as AGN activity and galaxy-galaxy interactions are unlikely to quench
on a large scale in A1795, and RPS would not strip the cold gas and
leave the more tenuously-bound hot halo.

\acknowledgements{}

CW was supported by an R. Samuel McLaughlin Fellowship and an Ontario
Graduate Scholarship, and SC acknowledges support of the Natural Sciences
and Engineering Research Council of Canada through a generous Discovery
grant. We appreciate the constructive comments and suggestions by
the anonymous referee and Michael Balogh. This research made use of:
NASA's Astrophysics Data System; Ned Wright's online cosmology calculator
\citep{wright2006}; Eric O. Lebigot's \textit{Uncertainties: a Python
package for calculations with uncertainties}; \textsc{topcat}, an
interactive graphical viewer and editor for tabular data \citep{taylor2005};
and \textsc{Montage}, which is funded by the National Science Foundation
under Grant Number ACI-1440620, and was previously funded by the National
Aeronautics and Space Administration's Earth Science Technology Office,
Computation Technologies Project, under Cooperative Agreement Number
NCC5-626 between NASA and the California Institute of Technology.

\bibliographystyle{aasjournal}
\bibliography{bibtex}

\end{document}